\def\simlt{\mathrel{\rlap{\lower 3pt\hbox{$\sim$}}\raise
2.0pt\hbox{$<$}}}
\def\simgt{\mathrel{\rlap{\lower 3pt\hbox{$\sim$}} \raise
2.0pt\hbox{$>$}}}

\def\Msun{M_{\odot}}

\newcommand{\q}{\begin{equation}}
\newcommand{\qa}{\begin{eqnarray}}
\newcommand{\qs}{\begin{eqnarray*}}
\newcommand{\nq}{\end{equation}}
\newcommand{\nqa}{\end{eqnarray}}
\newcommand{\nqs}{\end{eqnarray*}}
\def\be{\begin{equation}}
\def\ee{\end{equation}}

\def\etal{{\it et al.~}}

\documentclass[useAMS,usenatbib,usegraphicx]{mn2e}
\usepackage{natbib}

\def\aap{A\&A }

\def\apj{ApJ }

\def\mnras{MNRAS }

\newcommand{\gsim}{\raisebox{-3.8pt}{$\;\stackrel{\textstyle >}{\sim}\;$}}

\title[Global star formation histories and the galaxy Populations]{Simulating 
galaxy Clusters --II: global star formation histories and the galaxy 
Populations}
\author[Romeo, Portinari, Sommer-Larsen]{
Alessio D. Romeo$^{1,2,3}$ \thanks{E-mail: aro@ct.astro.it}, 
Laura Portinari$^{1,4}$\thanks{E-mail: lporti@utu.fi},
Jesper Sommer-Larsen$^{1,3,5}$\thanks{E-mail: jslarsen@tac.dk},\\
$^{1}$ Theoretical Astrophysics Center, Juliane Maries Vej 30, 2100 
Copenhagen \O, Denmark\\
$^{2}$ Dipartimento di Fisica e Astronomia, Universit\`{a} di Catania, 
via S.Sofia 64, 95123 Catania, Italy\\
$^{3}$ Nordita, Blegdamsvej 17, 2100 Copenhagen \O, Denmark\\
$^{4}$ Tuorla Observatory, V\"ais\"al\"antie 20, FIN-21500 Piikki\"o, 
Finland\\
$^{5}$ Astronomical Observatory, University of Copenhagen, 
Juliane Maries Vej 30, 2100 Copenhagen \O, Denmark}
\begin{document}

\date{Accepted ?. Received ?; in original form ?}

\pagerange{\pageref{firstpage}--\pageref{lastpage}} \pubyear{2004}

\maketitle

\label{firstpage}
\begin{abstract}
We performed N--body + hydrodynamical simulations of the formation and
evolution
of galaxy groups and clusters in a $\Lambda$CDM cosmology.
The simulations invoke star formation, chemical evolution with
non-instantaneous recycling, metal dependent radiative cooling, strong 
starbursts and (optionally) AGN driven galactic
super winds, effects of a meta-galactic UV field and thermal conduction.
The properties of the galaxy populations in two clusters, one Virgo-like
($T$$\sim$ 3 keV) and one (sub) Coma-like ($T$$\sim$6 keV), are discussed.
The global star formation rates of the cluster galaxies are found to 
decrease very significantly from redshift $z$=2 to 0, in agreement
with observations. The total K-band luminosity of the cluster galaxies 
correlates tightly with total cluster mass, and for models without additional
AGN feedback, the zero point of the relation matches the observed one
fairly well.
Compared to the observed galaxy luminosity function, the simulations
nicely match the number of intermediate--mass galaxies 
{\mbox{(--20$\la M_{\rm{B}} \la$--17,}}
smaller galaxies being affected by resolution limits) but they show
a deficiency of bright galaxies in favour of an overgrown central cD. High
resolution tests indicate that this deficiency is {\it not} simply due 
to numerical ``over--merging''.

The redshift evolution of the luminosity functions from $z$=1 to 0 is mainly 
driven by luminosity evolution, but also by merging of bright galaxies 
with the cD. 

The colour--magnitude  
relation of the cluster galaxies matches the observed ``red sequence'',
though with a large scatter,
and on average galaxy metallicity increases with luminosity. As the
brighter galaxies are essentially coeval, the colour--magnitude relation 
results from metallicity rather than age effects, as observed. On the whole,
a top-heavy IMF appears to be preferably required to reproduce also 
the observed colours and metallicities of the stellar populations.
\end{abstract}

\begin{keywords}
cosmology: theory --- cosmology: numerical simulations --- galaxies: clusters 
--- galaxies: formation --- galaxies: evolution 
\end{keywords}

\section{Introduction}
Clusters of galaxies are of great interest both as cosmological probes and
as ``laboratories'' for studying galaxy formation. The mass function and 
number density of galaxy clusters as a function of redshift is a powerful 
diagnostic for the determination of  
cosmological parameters (see Voit 2004 for a recent comprehensive review). 
Besides, clusters represent 
higher than average concentrations of galaxies, 
with active interaction and
exchange of material between them and their environment, as testified by
the non-primordial composition of the hot surrounding gas (e.g.\
Matteucci \& Vettolani 1988; 
Arnaud {\it et~al.}\ 1992; Renzini {\it et~al.}\ 1993; 
Finoguenov {\it et~al.}\ 2000, 2001; De Grandi {\it et~al.}\ 2004;
Baumgartner {\it et~al.}\ 2005). 

There are observational and theoretical arguments indicating that clusters 
are not fair samples of the average global properties of the Universe: the 
morphological mixture of galaxies in clusters is significantly skewed toward
earlier types with respect to the field population, implying
star formation histories peaking at higher redshifts than is typical in the 
field (Dressler 1980; Goto {\it et~al.}\ 2003; Kodama \& Bower 2001; 
see also Section~3); this is qualitatively in line with the expectation
that high density regions 
such as clusters, in a hierarchical bottom--up cosmological scenario, evolve 
at an ``accelerated'' pace with respect to the rest of the Universe
(Bower 1991; Diaferio {\it et~al.}\ 2001; Benson {\it et~al.}\ 2001).
Although clusters are somewhat biased sites of galaxy formation, 
they present the advantage of being bound structures with 
deep potential wells, likely to retain all the
matter that falls within their gravitational influence; henceforth, they
represent well-defined, self--contained ``pools'', where one can aim at keeping
full track of the process of galaxy formation and evolution, and of the
global interplay between galaxies and their environment.

The physics of clusters of galaxies has thus received increasing attention
in the past decade, benefiting from a number of X--ray missions measuring
the emission of the hot intra--cluster gas (e.g., {\it ASCA, 
ROSAT, XMM, Chandra}) as well as from extensive optical/NIR surveys probing
the distribution of galaxies and their properties (e.g.\ MORPHS, SDSS, 2MASS). 
Understanding cluster physics is also crucial to reconstruct, from the observed
X--ray luminosity function and temperature distribution, the intrinsic
mass function of clusters as a function of redshift, which is a quantity
of profound cosmological interest (Voit 2004).

The baryonic mass in clusters is largely in the form of a hot intra--cluster 
medium (ICM), which dominates by a factor of 5--10 over the stellar mass 
(Arnaud {\it et~al.}\ 1992; Lin, Mohr \& Stanford 2003). Consequently, 
early theoretical work and numerical simulations concentrated on pure 
gas dynamics when modelling
clusters. Recently however, attention has focused also on the role of galaxy
and star formation, and related effects. Star formation locks--up low
entropy gas, and supplies thermal and kinetic energy to the surrounding medium
via supernova explosion and shell expansion; both processes likely contribute
to the observed ``entropy floor'' in low--mass clusters, and the corresponding
breaking of the scaling relations expected from pure gravitational collapse
physics (Voit {\it et~al.}\ 2003). Besides, star formation is 
accompanied by the production of new metals
and the chemical enrichment of the environment; the considerable (about 1/3 
solar) 
metallicity of the ICM
indicates that a significant
fraction of the metals produced --- comparable or even larger than 
the fraction remaining within the galaxies --- is dispersed into the 
intergalactic medium (Renzini 2004), affecting the cooling rates of the
intra--cluster
gas. It is thus clear that the hydrodynamical evolution
of the hot ICM is intimately connected to the formation and evolution of 
cluster galaxies.
Only recently, however, due to advances in computing capabilities
as well as in detailed physical modelling, cluster
simulations have reached a level of sophistication adequate to trace
star formation and related effects in individual galaxies, and the chemical
enrichment of the ICM by galactic winds, in a reasonably realistic way
(Valdarnini 2003; Tornatore {\it et~al.}\ 2004).

Indeed so far theoretical predictions of the properties of cluster galaxy
populations within a fully cosmological context, have been mainly derived
by means of semi--analytical models. High resolution, purely N--body 
cosmological simulations of the evolution of the collisionless dark matter 
component, are combined with semi--analytical recipes describing
galaxy formation and related physics (such as chemical enrichment, stellar
feedback and exchange of gas and metals between galaxies and their
environment; see e.g.\  De Lucia, Kauffmann \& White 2004 for a recent
reference); with such schemes, the evolution of galaxies is ``painted'' on top 
of that of the simulated dark matter haloes and sub--haloes. The advantage
of this technique is that very high resolution can be attained, since pure 
N--body simulations can handle larger numbers of particles than
hydrodynamical simulations; besides, a wide
range of parameters can be explored for baryonic physics (e.g.\ star formation 
and feedback efficiency, Initial Mass Function, etc.).

In this paper, we present for the first time (to our knowledge) an analysis
of the properties of the galaxy population of clusters as predicted directly
from cosmological simulations including detailed baryonic 
physics, gas dynamics and galaxy formation and evolution. The resolution
for N--body + hydrodynamical simulations cannot reach the level of the
purely N--body simulations that constitute the backbone of semi--analytical
models, so we cannot resolve galaxies at the faint end 
of the luminosity function ($M_B$$\ga$--16). 
On the other hand, our simulations have
the advantage of describing 
the actual hydrodynamical response of the ICM to star formation, stellar
feedback and chemical enrichment. Although some uncertain physical
processes still necessarily rely on parameters (like the star formation 
efficiency and the feedback strength, see Section~2), once these are
chosen, the interplay between cluster galaxies 
and their environment follows
in a realistic fashion, as part of the global cosmological evolution 
of the cluster.

Moreover, the intimate relation between stellar Initial Mass Function (IMF), 
stellar luminosity, chemical enrichment, supernova energy input, returned 
gas fraction and gas flows out of/into the galaxies is included 
self--consistently in the simulations (while 
sometimes these are treated as adjustable, {\it independent} parameters in 
semi--analytical models).
The properties we obtain for the galaxy populations
(notably, global star formation rates, luminosity functions and 
colour--magnitude relations) are then
the end result of {\it ab--initio} simulations, with a far minor degree of
parameter calibration than in semi--analytical schemes.

In a standard $\Lambda$CDM cosmology, we have performed N-body +
 hydrodynamical (SPH) simulations of
the formation and evolution of clusters of different mass, on scales 
of groups to moderately rich clusters (emission-weighted temperature from 
1 to 6~keV).
In Paper~I of this series (Romeo {\it et~al.}\ 2004, in preparation) we
analyze the properties and distribution of the hot ICM in the simulated 
clusters, and discuss the effects that star formation and related baryonic 
physics have, on the predicted X--ray properties of the hot gas.
Several sets of simulations have been carried out, assuming different IMFs and 
feedback prescriptions (see Paper~I and Section~2). The chemical and 
X--ray properties of the ICM are best reproduced assuming a fairly top--heavy
IMF and a high, though not extreme, feedback (super-wind) efficiency (the
simulations marked, hereafter, as AY-SW; see Paper~I and Table 1).

In this Paper~II we focus instead on the properties of the galaxy population
in our simulated richer clusters (with temperatures between 3 and 6~keV),
where the number of identified galaxies is statistically significant.
We will mainly discuss the results from the AY-SW simulations, favoured by 
the resulting properties of the ICM; results from simulations with different 
input physics (IMF, wind efficiency, preheating) are also discussed for 
comparison, where relevant.

In Section~2 we briefly introduce the code and the simulations (full details
are given in Paper~I), as well as the procedure to identify cluster galaxies
in the simulations. 
In Section~3 we discuss the global star formation histories
of cluster galaxies, and in Section~4 we determine global luminosities of the
clusters simulated with different prescriptions for baryonic physics. 
In Section~5 and~6 we discuss luminosity functions and colour--magnitude
relations of the galaxy population in our clusters, and, finally, in 
Section~7 we summarize our results. 

\section{The simulations}
\label{sect:simulations}

The code used for the simulations is a significantly improved version of
the TreeSPH code we used previously for galaxy formation simulations 
(Sommer-Larsen, G\"otz \& Portinari 2003). 
Full details on the code and the simulations are given in Paper~I, here we
recall the main upgrades over the previous version.
(1) In lower resolution regions an improvement in the numerical
accuracy of the integration of the basic equations
is obtained by incorporating the ``conservative'' 
entropy equation solving scheme of Springel \& Hernquist (2002).
(2) Cold high-density gas is turned into stars in a probabilistic way as
described in Sommer-Larsen \etal (2003). In a star-formation event 
a SPH particle
is converted fully into a star particle. Non-instantaneous recycling of
gas and heavy elements is described through probabilistic ``decay'' of star 
particles back to SPH particles as discussed by Lia et~al.\ (2002a). 
In a decay event a 
star particle is converted fully into a SPH particle.
(3) Non-instantaneous chemical evolution tracing
10 elements (H, He, C, N, O, Mg, Si, S, Ca and Fe) has been incorporated
in the code following Lia et~al.\ (2002a,b); the algorithm includes 
supernov\ae\ of type II and type Ia, and mass loss from stars of all masses.
For the simulations presented in this paper, a Salpeter (1955) IMF and
an Arimoto--Yoshii (1987) IMF were adopted, both with 
with mass limits [0.1--100]~$\Msun$.
(4) Atomic radiative cooling is implemented, depending both on the 
metallicity of the gas (Sutherland \& Dopita 1993) and on the meta--galactic 
UV field, modelled after Haardt \& Madau (1996). 
(5) Star-burst driven, galactic super-winds are incorporated in the 
simulations. This is required to expel metals from the galaxies and reproduce
the observed levels of chemical enrichment of the ICM.
A burst of star formation is modelled in the same way as the ``early 
bursts'' of Sommer--Larsen {\it et~al.}\ (2003): the energy released 
by SNII explosions goes initially into the interstellar medium as thermal 
energy,
and gas cooling is locally halted to reproduce the adiabatic super--shell
expansion phase; a fraction of the supplied energy is 
subsequently converted (by the hydro code itself) 
into kinetic energy of the resulting expanding super-winds and/or shells.
The super--shell expansion also drives the dispersion of the metals produced
by type~II supernov\ae\ (while metals produced on longer timescales
are restituted to the
gaseous phase by the ``decay'' of the corresponding star particles, see point
2 above).
The strength of the super-winds is modelled
via 
a free parameter $f_{\rm{wind}}$ which determines how large a fraction
of the new--born stars partake in such bursting, super-wind 
driving star formation. We find that
in order to get an iron abundance in the ICM comparable to observations,
$f_{\rm{wind}}\ga 0.5$ and at a top-heavy IMF 
must be used. 
(6) Thermal conduction was implemented in the code following Cleary \&
Monaghan (1999), with the addition that 
effects of saturation 
(Cowie \& McKee 1977) were taken into account.

The groups and clusters were drawn and re-simulated from a dark matter 
(DM)-only cosmological simulation run with the code FLY
(Antonuccio {\it et al.}, 2003), for a standard flat $\Lambda$ Cold Dark 
Matter cosmological model ($h=0.7$, $\Omega_0=0.3$, 
$\sigma_8=0.9$) with $150 h^{-1}$~Mpc box-length. 
When re-simulating with the hydro-code, baryonic 
particles were ``added" to the original DM ones, which were
split according to a chosen baryon fraction $f_b=0.12$. 
This results in particle masses of $m_{\rm{gas}}$=$m_*$=
$2.5\cdot 10^8$ and $m_{\rm{DM}}$=$1.8\cdot 10^9$ $h^{-1}$M$_{\odot}$.
Gravitational (spline) 
softening lengths of $\epsilon_{\rm{gas}}$=$\epsilon_*$=2.8 and 
$\epsilon_{\rm{DM}}$=5.4 $h^{-1}$kpc, respectively, were adopted.
The gravity softening lengths were fixed in physical coordinates from $z$=6
to $z$=0 and in comoving coordinates at earlier times.
Particle numbers are in the range $4\cdot 10^5 - 10^6$ SPH+DM particles.
To test for numerical resolution effects one simulation was run with eight 
times higher mass and two times higher force resolution, with $2.3\cdot 10^6$ 
SPH+DM particles having $m_{\rm{gas}}$=$m_*$=
$3.1\cdot 10^7$, $m_{\rm{DM}}$=$2.3\cdot 10^8$ $h^{-1}$M$_{\odot}$ and 
softening lengths of 1.4, 1.4 and 2.7 $h^{-1}$kpc, respectively
(Section~\ref{sect:resolution}).

We selected from the cosmological simulation 
some fairly relaxed systems, not undergoing any major mergers since $z \la 1$: 
two galaxy groups, two small mass clusters, one ``Virgo-like'' cluster
($T\sim 3$ keV) and a ``Mini-Coma'' ($T\sim 6$ keV) cluster. 
TreeSPH re-simulations were run with different super-wind (SW)
prescriptions, IMFs, either Salpeter (Sal) or Arimoto--Yoshii (AY), 
with or without thermal conduction, with or without 
preheating and, as mentioned above, at two different numerical resolutions.

As our reference runs we shall consider simulations with an AY IMF, 
with 70\% of the energy feedback from supernovae type II going into driving 
galactic super-winds, 
zero conductivity, and at normal resolution; 
these will be denoted ``AY-SW''. Some other simulations 
were run with galactic super-wind feedback two times (SWx2) and four
times (SWx4) as energetic as is available from supernovae; the additional
amount of energy is assumed to come from AGN activity. Others were of the
AY-SW type, either with additional preheating at $z$=3
(preh.), as discussed by Tornatore {\it et~al.}\ (2003), or with thermal 
conduction included (COND), assuming a conductivity of 1/3 of the 
Spitzer value (e.g., Jubelgas, Springel \& Dolag 2004).
Finally, one series of
simulations was run with a Salpeter IMF and only early ($z$$\ga$4), strong 
feedback, as in the galaxy formation simulations of  Sommer-Larsen 
{\it et~al.}\ (2003); as this results in overall fairly weak
feedback we denote this as: Sal-WFB. We refer the reader to Papers I and III,
and also Sommer-Larsen (2004), for more details. 

The general features and main results for all the simulations are
presented in the companion Paper~I. In the present paper
we mainly discuss the properties of the galaxy populations 
in the simulated ``Virgo'' and ``Mini-Coma'' clusters, which are 
in Paper~III referred to as clusters ``C1'' and ``C2'', respectively.
We focus on these two largest simulated objects since they contain
a significant number of identified individual galaxies.
The prescriptions adopted for the re-simulations of these two clusters
are summarized in Table~1.
Lower mass objects
are only included in the analysis of the luminosity scaling in Section~4. 

We consider the AY-SW simulation as our ``standard'' run. 
These
simulations provide a satisfactory overall match to the observed properties 
of the ICM (see Paper~I, and also Sommer-Larsen 2004).
For ``Virgo'' we also have simulations with higher feedback efficiency,
with pre-heating or with the Salpeter IMF (weak or normal feedback); the
corresponding results for the galaxy population are shown for comparison,
where relevant (e.g., Figs.~\ref{fig:LFV} and~\ref{fig:rsVirgo}).

Because of its high
computational demand, the ``Coma'' system was run only with the standard AY-SW 
prescription, with (COND) or without thermal conduction.

\begin{table*}
\caption{Basic physical properties and prescriptions adopted for the runs 
of the ``Virgo'' and ``Coma'' clusters discussed in this paper. 
Temperatures ($4^{th}$ column) refer to the
``standard'' run AY-SW. The last two columns give quantities as resulted
from the simulations, at $R_{500}$.}
\begin{tabular}{c c c c c c c c c c}
\hline
cluster &       $M_{vir}$       & $R_{vir}$ & $<k T_{ew}>$ & IMF &  SN feedback  
& pre-heat. @ z=3 &   therm. cond.  & $f_{cold}$ & $\frac{Z_{Fe}^{ICM}}
{Z_{Fe,\odot}}$ \\
        & [$10^{14}~M_{\odot}$] &   [Mpc]   &    [keV]     &     & efficiency 
&  [keV/part.]  & (Spitzer)  &   ($R_{500}$) &       \\
\hline
``Virgo'' & 2.77 & 1.62 & 3.0 & AY & SW & --- & --- & 0.19 & 0.21 \\ 
                                                                  
          & & & & AY & SW & --- & 1/3 & 0.17 & 0.38 \\	
\          & & & & AY & SW$\times$ 2 & ---  & --- & 0.11 & 0.22 \\	
          & & & & AY & SW$\times$ 4 & --- & --- & 0.06& 0.16 \\	
	  & & & & AY & SW & 0.75  & --- & 0.16 & 0.22 \\	
                                                                    
          & & & & AY & SW & 1.5  & ---& 0.15 & 0.30 \\	
	                                                             
          & & & & Sal.        & SW & --- & --- & 0.23 & 0.14 \\  
          & & & & Sal.        & Weak  & --- & --- & 0.27 & 0.12 \\  
\hline
``Coma'' & 12.40 & 2.90 & 6.0 & AY & SW & --- & --- & 0.20 & 0.20 \\ 
                                                                     
         & & & & AY & SW & --- & 1/3  & 0.17 & 0.20 \\	
\hline
\end{tabular}
\label{tab:prescriptions}
\end{table*}
\subsection{The identification of individual galaxies}

The stellar contents of both clusters are characterized by a massive, 
central dominant (cD) elliptical galaxy surrounded by other galaxies orbiting
in the main cluster potential, and embedded
in an extended envelope of tidally stripped intra-cluster stars, unbound from 
the galaxies themselves. Observationally, the brightest galaxy in clusters 
is usually located at or near the cluster centre, where other (massive) 
galaxies likely
tend to sink and merge by dynamical friction, mainly against the dark matter. 
Our simulated cDs appear very dominant, 
containing slightly more than half of all the star particles in the cluster, 
in this being more prominent 
than the cDs observed in real clusters. This is related to a 
``central cooling'' problem common in models and hydrodynamical simulations,
both SPH and Adaptive Mesh Refinement (e.g.\
Ciotti et~al.\ 1991; 
Fabian 1994; 
Knight \& Ponman 1997; Suginohara \& Ostriker 1998; Lewis et~al.\ 2000; 
Tornatore et~al.\ 2003; Nagai \& Kravtsov 2004): 
after the cluster has recovered from the last major merging events, 
a steady cooling flow is established at the center of 
the cluster, being only partially attenuated by the strong feedback. The
cooled-out gas is turned into stars at the base of the cooling flow 
($r\la 10$ kpc). This fairly young stellar population, accumulating at
the centre of the cD, contributes very significantly to the total luminosity, 
and increases the total stellar mass of the cD by of the order a factor of
two. In semi--analytical models, one can 
alleviate this problem by artificially quenching radiative gas cooling in
galactic haloes more massive than 350~km/sec (Kauffmann {\it et~al.}\ 1999).
However, over--cooling in cluster cores is not merely a technical 
problem in numerical simulations, but a problem with the physics in the 
central cluster regions. {\it XMM} and {\it Chandra} observations have revealed
that cooling occurs only down to temperatures of about 1--2~keV, so that
the former ``cooling flow'' regions are now known to be instead ``cool cores''
(e.g.\  Molendi \& Pizzolato 2001; Peterson et~al.\ 2001, 2003; 
Tamura et~al.\ 2001; Matsushita et~al.\ 2002);
the mechanisms that prevent the gas from cooling further and finally 
form stars at a high rate, are presently not clearly identified (central AGN 
feedback, magnetohydrodynamic effects, thermal conduction and/or heating 
through gravitational interactions are some candidates; e.g.\
Ciotti \& Ostriker 2001; Narayan \& Medvedev 2001; Makishima et~al.\ 2001; 
Fabian, Voigt \& Morris 2002; Churazov et~al.\ 2002; Kaiser \& Binney 2003; 
Fujita, Suzuki \& Wada 2004; Cen 2005).

The cD and its envelope of intracluster stars formed in our simulations
are discussed in the companion Paper~III of this series (Sommer-Larsen, 
Romeo \& Portinari, 2005). 
Here, we study the properties of the rest of the galaxy population in the
simulated clusters.
The galaxies are identified in the simulations by means of the procedure 
detailed in Paper~III.
Visual inspection of the z=0 frames shows that the stars in all
galaxies (except the cD) are typically located within 10--15 kpc from the 
respective galactic centers. In a cubic grid of cube--length 10~kpc, we
identify all cells containing at least 2 star particles. Each of these is then
embedded in a larger cube of length 30~kpc; if this larger cube contains a 
minimum of $N_{min}=$7 gravitationally bound star particles, the system is 
labelled as a potential galaxy. Finally, among the various potential galaxies 
effectively identifying the same system, we classify as {\it the} galaxy the 
one containing the largest number of star particles. 
With the mass resolution of the simulations, $N_{min}$ corresponds 
to a stellar mass of $2.5 \times 10^9 \, M_{\odot}$ and an absolute B-band
magnitude of $M_B \approx -16$, of the order of that of the Large Magellanic 
Cloud. 
The galaxy identification algorithm is adequately robust, as long as 
$N_{min}$$\sim$7-10 star particles. Note though that 
galaxies will contain gas and dark matter particles as well, in the case
of small galaxies often considerably more dark matter particles, because
most of the baryons have been expelled by galactic super-winds. 

Using this galaxy identification procedure, we identify for the standard
runs 42 galaxies in ``Virgo'' and 212 in ``Coma'', respectively. 

\subsection{The computation of the luminosity}
\label{sect:lumcomp}

We assign luminosities and colours to the galaxies 
identified in our simulations, as the sum of the luminosities 
of the relevant star particles in the various passbands.
 
Each star particle represents 
a Single Stellar Population (SSP) of total mass of
$3.6 \times 10^8 \, M_{\odot}$, with individual stellar masses distributed
according to a particular IMF (either Arimoto-Yoshii or Salpeter), and we 
keep record of the age and the metallicity 
of each of these SSPs. It is quite straightforward to compute
the global luminosities and colours of our simulated galaxies, as the sum
of the contribution of their constituent star particles. SSP luminosities
are computed by mass-weighted integration of the Padova 
isochrones (Girardi {\it et al.}, 2002).

We must however pay special attention to the relation between star particle
mass and SSP mass.
In our ``statistical'' implementation of chemical evolution (Lia {\it et~al.}\
2002a,b), a fraction of the star particles with age transforms back
into gas particles (the gas--again particles of Lia {\it et~al.}), to simulate
the re-ejection of gas into the interstellar medium by dying stars. Following
the notation by Lia {\it et~al.}, the re--ejected fraction increases with the 
age $t$ of the SSP:
\[ E(t) = \int_{M(t)}^{M_u} \frac{M-M_r(M)}{M} \, \Phi(M) \, dM \]
where $\Phi(M)$ is the IMF, $M_u$ is its upper mass limit, and $M(t)$ is the 
mass of the star with lifetime $t$. Typical values of the global returned 
fraction after a Hubble time are 30\% for the Salpeter IMF, and around 50\%
for the Arimoto--Yoshii IMF.
As a consequence, out of an episode of star formation involving $N$ star
particles, after a time $t$ on average only $N\times (1-E(t))$ remain, 
while $N\times E(t)$ have returned to be SPH particles.
We need to take this effect into account when computing the luminosities,
since SSP luminosities refer to the {\it initial} mass of the SSP, namely 
the mass involved in the original star formation episode, not to its current
mass which is a fraction $1-E(t)$ of the initial one. Each star particle 
of age $t$ is effectively representative of $\frac{1}{1-E(t)}$ star
particles at its birth 
--- or equivalently, each star particle of mass $m_*$ corresponds to an
initial SSP mass $\frac{m_*}{1-E(t)}$. Therefore, for the computation of the
luminosity each star particle is assigned a corresponding ``initial SSP mass''
$\frac{m_*}{1-E(t)}$, rather than its actual present mass $m_*$. 
\section{Star formation histories of the cluster galaxies}
\label{sect:SFH}

The star formation process in the cluster environment is known to peak 
at higher redshifts than in the field.
The morphology--density relation (e.g.\
Dressler 1980; 
Goto {\it et~al.} 2003) implies that the cluster population is
dominated by early--type galaxies, ellipticals and S0's, whose stellar 
populations have formed at redshift $z >2$ as indicated by the 
tightness and the redshift evolution of the colour--magnitude relation and 
of the fundamental plane (e.g.\
Bower, Lucey \& Ellis 1992; 
Kodama \& Arimoto 1997; 
J{\o}rgensen {\it et~al.} 1999; 
van Dokkum \& Stanford 2003). 
Conversely, the field galaxy population is dominated by late Hubble types, 
with star formation histories still presently on--going.
Furthermore, compared to their cluster counterparts,
field ellipticals 
(especially low mass ones) display
more extended star formation histories and/or minor
star formation episodes at low redshifts (e.g.\ Bressan {\it et~al.} 1996; 
Trager {\it et~al.} 2000; 
Bernardi {\it et~al.} 2003; 
Treu {\it et~al.} 2005).
The rapid drop in the past star formation rate in clusters 
is also apparent in the evolution of the fraction of star
forming galaxies, as traced by blue colours and spectroscopy 
(e.g.\ Butcher \& Oemler 1978, 1984;
Ellingson {\it et~al.} 2001; Margoniner {\it et~al.} 2001; 
Poggianti {\it et~al.} 1999, 2004; 
Gomez {\it et~al.} 2003). 

These trends are qualitatively in line with present theories of cosmological 
structure formation, predicting that high density regions evolve faster
than low--density regions\footnote{It is worth mentioning that even 
when modelling isolated galaxy evolution (in the monolithic collapse fashion), 
a more prolongued 
star formation history is predicted in low--mass,
low density galaxies, with respect to the more massive or denser 
ones (Carraro {\it et~al.}\ 2001; Chiosi \& Carraro 2002).}
(Bower 1991; Diaferio {\it et~al.} 2001; Benson 
{\it et~al.} 2001); although there are severe quantitative discrepancies with 
the results of semi--analytical hierarchical models, especially concerning 
the formation and number density evolution of field early--type galaxies 
(van Dokkum {\it et~al.} 2001; Benson {\it et~al.} 2002).

All in all,
clusters
display a drastic decay in the star formation rate at $z<1$, much faster 
than the corresponding decline indicated by the Madau plot for field galaxies 
(Kodama \& Bower 2001; Fig.~1). 
%
Such drop in the cluster star formation rate 
is usually ascribed to a combination of three effects. 
(1) Interaction with the cluster environment and with the ICM quenches 
the star formation of the infalling galaxies 
via 
mergings and interactions, 
harassment, ram pressure stripping, 
starvation/strangulation, 
etc.\ (Dressler 2004 and references therein).
(2) The rate of accretion of 
star forming field galaxies onto clusters decreases at decreasing redshift
(Bower 1991; Kauffmann 1995). 
(3) The intrinsic star formation rate of the accreted 
galaxies 
decreases at $z < 1$ (cf.\ the Madau plot, Madau {\it et~al.} 
1998).

In this section, we analyze the global star formation history of the cluster
galaxies in our simulations and compare it to the observational one 
derived by Kodama \& Bower (2001). Just like for the empirical one, our
global star formation history is obtained as the sum of the star formation 
histories of the individual cluster galaxies 
(excluding the young cD stars in the central regions, formed out of
the cooling flow; see also Paper~III).

For each cluster, we single out all the star particles that at the
present time ($z$=0) belong to the 
identified galaxies within 1~Mpc,
corresponding to the same $R_{30}$ radius considered by Kodama \& Bower.
For each star particle we know the age $t$ and we 
derive the corresponding ``initial SSP mass'' involved in star formation
at its birth time, as $\frac{m_*}{1-E(t)}$ (see \S~\ref{sect:lumcomp}).
Summing over all the selected star particles, we reconstruct
the past SFR in the comoving Lagrangian volume.

\begin{figure}
\includegraphics[width=70mm]{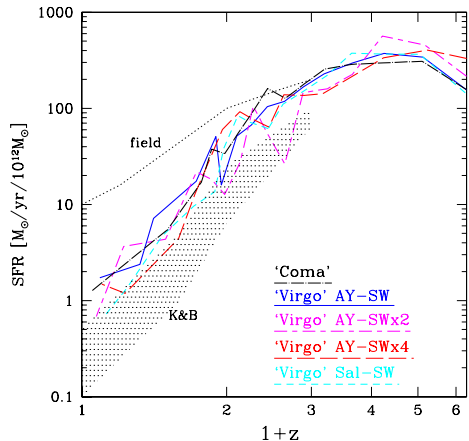}
\caption{Normalized global star 
formation rates in the ``Virgo'' and ``Coma'' galaxies, excluding 
the youngest stars (less than 1~Gyr old) in the 
inner 10 kpc of the cDs; symbols are explained in section~2. Also shown 
are observed rates for field galaxies ({\it dotted}, Madau {\it et al.}, 1998) 
and those of galaxies in the inner parts of rich clusters ({\it shaded} 
region, from Kodama \& Bower, 2001).}
\label{fig:SFRz}
\end{figure}

In Fig.~1 the global normalized SFR is shown for the galaxies in ``Virgo'' 
(3 models with varying feedback strength and Arimoto--Yoshii IMF, 
plus a model with Salpeter IMF) and ``Coma'' (standard reference
prescription). 
Since $z\sim 1.2$ the SFR declines considerably more rapidly than that of the 
field galaxies,
in line with observational reconstructions
indicating
a peak at $z\sim 3-4$ and a subsequent significative drop from $z\simeq 2$ 
to 0 quite more pronounced than in the field.
In Fig.~\ref{fig:SFRz}, the shape of the SFR history in our simulated
clusters is in agreement with the estimates of Kodama \& Bower, however the
normalization is somewhat higher (close to the field level at redshift $z=2$).

From comparing ``Virgo'' and ``Coma'' in the figure, 
it is evident that the total star formation history  normalized by
cluster mass is nearly independent of the cluster virial mass, as also
recently pointed out by Goto (2005) over a sample of 115 nearby clusters
selected from the SDSS: this 
suggests that physical mechanisms depending on virial mass 
(such as ram-pressure stripping) are not exclusively driving 
galaxy evolution within clusters.

\section{Scaling properties of the cluster light}
\begin{figure}
\includegraphics[width=73mm]{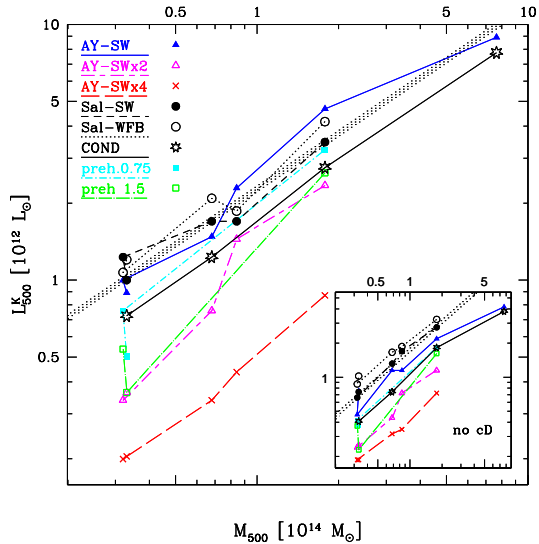}
  \caption{{\it K}-band luminosity-mass relation inside of $R_{500}$ at $z$=0
(excluding cD stars in the innermost 10 kpc, younger than 1 Gyr); 
notice that ``Coma'' cluster results at $M_{500} \sim 7 \times 
10^{14}~M_{\odot}$ are available only for the AY-SW prescription, with and 
without thermal conduction (see Table~1).
{\it Shaded strip}: best-fit relation for a sample of 93 clusters and groups 
from 2MASS (Lin {\it et al.}, 2004). In the insert: the same excluding all 
stars born since
 $z\simeq 1$ from the central 40 kpc; 
also the observational best-fit relation has been modified to exclude the BCG.}
\label{LM}
\end{figure}

The near-infrared luminosity of galaxies is only negligibly affected by
recent star formation activity, thus giving 
a robust measure of the actual stellar content of a cluster
(Kauffmann \& Charlot 1998).
In Fig.~\ref{LM} we show the total (2.2$\mu$) K-band luminosity of galaxies 
within $r_{500}$ versus the total mass inside of $r_{500}$ (which in the 
$\Lambda$CDM model is approximately half of the virial
radius) for individual clusters in the simulated sample --- including 
objects of lower mass than ``Virgo'' and ``Coma''.
The various models match quite well the 
best-fit slope derived observationally by Lin, Mohr \& Stanford 2004:
$L^K_{500}\propto M_{500}^{0.72}$, but the
runs with additional pre-heating and those with stronger feedback result 
in a normalization too low with respect to the observations. 
As to the Salpeter simulations, the excellent agreement seen
in this plot is partly affected by the very blue colours of the galaxies
(Fig.~\ref{fig:rsVirgo}); in bluer bands, e.g.\ the B band, these
simulations results too bright for their cluster mass (Paper~I).

The figure's insert shows the result of correcting for the excess of young cD 
stars by neglecting the
luminosity contribution of stars in the innermost 40~kpc formed at 
$z \la 1$ (see Paper~III for details); the observational data have been 
corrected as well by excluding the cluster brightest (usually central) galaxy 
(BCG). The slope of the 
relation remains essentially the same, though observationally it steepens 
slightly to
$L^K_{500}\propto M_{500}^{0.83}$ (Lin {\it et~al.} 2004). 
This indicates  that the relative contribution of the cD/BCG to the total 
cluster light decreases with cluster mass. 

 The regular trend seen in Fig.~\ref{LM} over a considerable range of mass
suggests that star/galaxy formation in groups and clusters 
tends to be approximately a scaling process; moreover 
the slope of the observational relation  
indicates that smaller groups produce stars at higher efficiency than larger 
clusters, provided that {\it K}-band light well traces stellar mass.

\begin{figure}
\includegraphics[width=73mm]{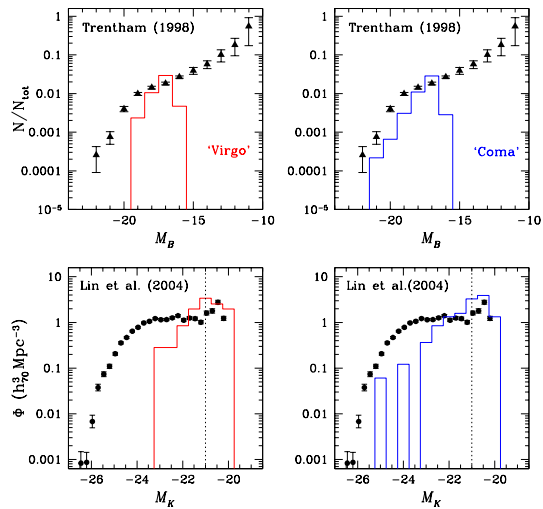}
  \caption{{\it Top panels}: B--band luminosity functions at $z$=0
for the standard (AY-SW) ``Virgo'' ({\it left}) and ``Coma'' ({\it right})
simulations; the LFs are normalized to have the same 
number of galaxies as the observed average LF (Trentham 1998, {\it triangles})
within the populated luminosity range ($-19 \leq M_B \leq -17$). 
{\it Bottom panels}:
K--band LFs for the same clusters, compared to the observed average LF by
Lin {\it et~al.}\ (2004); the dotted line marks the limit $M_K \leq -21$ where 
the observational estimate is considered reliable.}
\label{fig:LF_rel}
\end{figure}

\section{The luminosity function of cluster galaxies}
\label{sect:lf}
In this section we discuss the luminosity function of the population of 
galaxies
in our simulated ``Virgo'' and ``Coma'' clusters (excluding the central cDs,
which are discussed in Paper~III). As mentioned above, we identify for the
standard runs 42 galaxies in ``Virgo'' and 212 galaxies in ``Coma'',
respectively, with a resolution--limited stellar
mass $ \ga 2.5 \times 10^9 \, M_{\odot}$, of the order that of the Large 
Magellanic Cloud. This limit corresponds to $M_B \sim -16$ at $z$=0 (Fig.~3).


In Fig.~\ref{fig:LF_rel} 
we compare the B--band and K--band luminosity function (LF) 
of our simulated cluster galaxies, to the observational LFs by Trentham (1998;
a weighted mean of 9 low--redshift clusters), and by Lin {\it et al.}
(2004; a composite LF of 93 clusters and groups from 2MASS data). 

For the B--band LF (top panels), the number of resolved individual galaxies 
quickly drops for objects fainter than $M_B \sim -17$, both in ``Virgo'' and in
``Coma'', due to the above mentioned resolution limits and identification 
procedure.
Notice that, although we are missing the dwarf galaxies that largely dominate
in number, we are able to describe the bulk of the stellar mass and of the
luminosity in clusters, which is dominated by galaxies around $L_*$ while
dwarfs fainter than $M_B \simeq$--17 give a negligible contribution (Cross 
{\it et~al.}\ 2001; Blanton {\it et~al.}\ 2004)
--- though in the
simulations, too much of the luminosity and stellar mass expected in
objects around $L_*$, is instead accumulated in the central cD (see below).
Our B--band LF in Fig.~\ref{fig:LF_rel} is normalized so that, in the 
luminosity bins that are significantly populated ($-19 \leq M_B \leq -17$), 
the overall 
number of objects is the same as for the observed relative LF (Trentham 1998);
in this magnitude range, the shape of the predicted and observed LF 
is directly comparable. 
The simulated LF is steeper than the observed one, namely we underestimate
the relative number of bright galaxies.
One reason for this may be our somewhat biased selection of the group and 
clusters sample, which consists of fairly relaxed 
systems, in which no significant merging takes place since $z\la 1$
(see Paper~III). This means that most of the massive galaxies have already 
merged with the central cD by dynamical friction - 
which is not always the case for real clusters (one likely, nearby 
example of an unrelaxed cluster is Virgo, e.g., Binggeli et al. 1987).

In the K--band (bottom panels of Fig.~\ref{fig:LF_rel}), our simulated LF 
reaches fainter magnitudes than the limit $M_K<-21$ (dotted line), where 
the observational LF is considered reliable (Lin {\it et~al.}\ 2004). 
Therefore, we normalize our LF to match the observed number of galaxies 
within the populated magnitude range and above the resolution limit 
($-23.2 \leq M_B \leq -21$). In the K--band, the lack of massive galaxies
in relative number is even more evident.

\begin{figure}
\includegraphics[width=80mm]{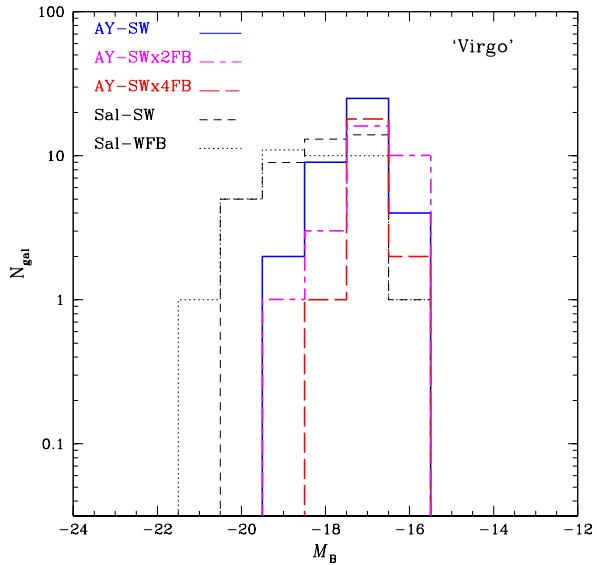}
  \caption{Same as in top panels of Fig.~\protect{\ref{fig:LF_rel}} 
for various ``Virgo'' simulations; symbols are explained in section~2.}
\label{fig:LFV}
\end{figure}

In Fig.~\ref{fig:LFV} we compare the LFs
for the ``Virgo'' cluster simulated with different IMFs and feedback 
prescriptions. For the three AY simulations, increasing the feedback strength
results in decreasing the masses of all galaxies, and hence, in particular,
the number of bright galaxies.
The results with thermal conduction (not shown in the figure)
are very close to the analogous case without thermal conduction.
The LF in the Salpeter case is broader in luminosity, due
the lower stellar feedback
with respect to the AY IMF, allowing a larger accumulation of stellar mass in 
galaxies. The trend is even stronger in the Sal-WFB case. 
Notice that, though the Salpeter IMF simulations are more successful 
in forming massive galaxies and would thus compare more favourably 
to the observed LF, this comes at the expense of a too large cold fraction 
with respect 
to observations (Table~\ref{tab:prescriptions} and Paper~I). Other reasons
to disfavour the Salpeter IMF simulations are the low predicted metallicities
in the ICM (Table~\ref{tab:prescriptions} and Paper~I), and the too blue
colours of the galaxies (\S~\ref{sect:RS}).

\begin{figure}
\includegraphics[width=90mm]{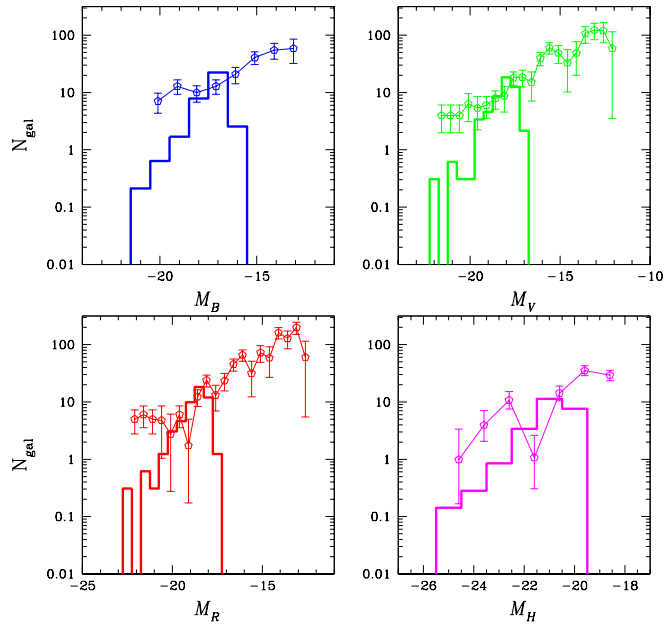}
  \caption{Absolute luminosity functions at $z$=0 of the ``Coma'' cluster 
within the relevant projected area (see text) in the 
{\it B, V, R, H} bands, compared with observed Coma LFs from Andreon \&
 Cuillandre 2002 ({\it B, V, R}), and Andreon \& Pell\`o 2000 ({\it H}).}
\label{fig:LF_Coma}
\end{figure}

In Fig.~\ref{fig:LF_Coma} we compare the LF of the ``mini-Coma'' cluster, 
in {\it absolute} number of galaxies, to the observed Coma LFs: 
besides discussing the distribution of the stellar mass in the
simulations (cD vs.\ bright galaxies vs.\ dwarf galaxies) we want to test 
if the {\it actual number} of the galaxies formed in the simulation 
is sensible. The Coma LF is from Andreon \& Cuillandre (2002) and 
Andreon \& Pell\'o (2000); for the comparison, we compute
the luminosity functions of galaxies in our simulated cluster within 
projected, off-center areas comparable to the areas covered by the 
observations --- about (0.8~Mpc)$^2$ in B, (1~Mpc)$^2$ in V, R and 
(0.5~Mpc)$^2$ in H. There is quite good 
agreement with the observed number of galaxies over the luminosity 
range that is significantly populated by the galaxies identified in the 
simulation.
However, as already noticed from the relative LF in Fig.~\ref{fig:LF_rel},
we are ``missing" the bright end tail of the LF, a problem which does not
seem 
to be cured by increased resolution (see \S\ref{sect:resolution}). 
These results on the absolute LF confirm that the problem is not that
we form too few galaxies in the simulations (whence one could statistically
``miss'' the fewer galaxies at the bright end of the LF), but that in fact
there is a lack of bright galaxies in favour of the overgrown cD.

\begin{figure}
\includegraphics[width=70mm]{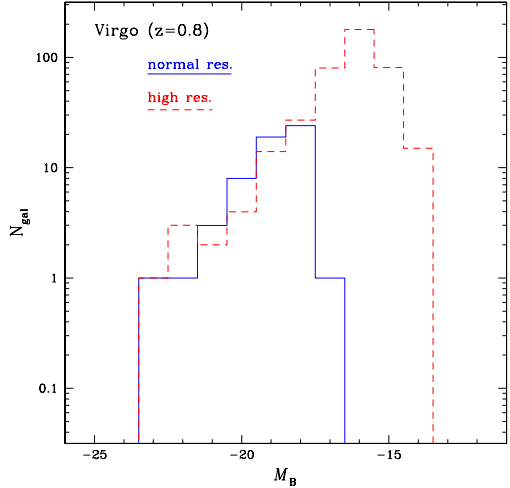}
  \caption{``Absolute'' luminosity function of ``Virgo" in the 
{\it B}-band, 
both at normal ({\it solid} histogram) and 8 times higher ({\it dashed}) 
resolution, at the epoch $z\simeq 0.8$.}
\label{fig:LFhrV}
\end{figure}

\subsection{Resolution effects}
\label{sect:resolution}

In order to test for numerical resolution effects, one simulation of the 
``Virgo'' system with the standard AY-SW model
was run at 8 times higher mass and 2 times higher force resolution. At
the time of writing, this simulation has reached $z\simeq$0.8, so that
in Fig.~\ref{fig:LFhrV} we can compare the Virgo LF for the two different
resolutions at $z \simeq$0.8.
As expected, at higher resolution the LF extends 
to fainter magnitudes. This is due to the higher force and mass resolution, 
which allows one to resolve as several small (proto-)galaxies an object which 
was previously initially identified as one large proto-galaxy.
However, above the resolution limit for the galaxy identification,
the LF appears to be strikingly robust to resolution effects, and consequently
so is all the discussion in the previous section. 
The global stellar content of the cluster is also 
essentially resolution independent: the total mass of stars at $z \simeq$0.8 
inside of the virial radius is $3.0\cdot 10^{12} M_{\odot}$ for the 
normal resolution case and $3.4\cdot 10^{12} M_{\odot}$ for the 
high-resolution one. The corresponding masses of stars in galaxies other
than the cD are 9.1 and 9.2$\cdot 10^{11} M_{\odot}$; the total masses of 
stars outside of the cD (adopting $r_{\rm{cD}}$=50~kpc) are 1.7 and 
2.0$\cdot 10^{12} M_{\odot}$, and the masses of stars in the cD are
1.3 and 1.4$\cdot 10^{12} M_{\odot}$, respectively.  
The total numbers of galaxies identified, assuming $N_{min}=$7 in both cases, 
are 50 and 341, respectively.

In conclusion, at higher resolution more substructure can be identified 
extending the LF at the dwarf galaxy end;
yet the luminosity of the second brightest galaxy (after the cD) 
in the two simulations of different resolution is quite similar. Moreover,
the number of galaxies brighter than $\approx L^*$ remains fairly
unaffected when going to higher resolution: enhancing
resolution does not increase the number of bright cluster galaxies. 
Analogous results are obtained from another resolution test we performed, 
on a smaller system (a group of galaxies of virial mass 
0.48 $\cdot 10^{14} M_{\odot}$) evolved down to $z=0$: 
the higher resolution does not affect the luminosity of the second ranked 
galaxy nor helps increasing the number of bright ($\sim L_*$) galaxies. 
These tests indicate that the lack of bright galaxies in the normal
resolution simulations of our "Virgo" and "Coma" clusters 
is {\it not} an effect of numerical
``over--merging''. 

``Classic" semi-analytical models of cluster formation also faced, 
to some extent, the problem that the cD became too bright and the number 
of lower ranked, bright galaxies was depleted. 
Springel {\it et~al.} (2001) suggested that numerical over--merging
onto the central cD in {\bf dissipationless dark matter simulations}
could deplete the number 
of galaxies at the bright end of the LF, and showed that for semi--analytical
models this problem can be significantly improved on by suitable
identification of dark matter substructure, or of ``haloes within haloes''.
This is not a solution in our case, however, because we 
identify as galaxies all bound systems containing as little as $N_{min}$=7 
star particles, and increasing resolution does not increase the number 
of bright galaxies. Moreover, in hydrodynamical simulations there is as of yet
no physical way to quench the central, semi-steady cooling flow; such late
gas accretion accounts for part of the excess of the cD masses.

\subsection{Redshift evolution of the Luminosity Function}

The LF in clusters can evolve due to two effects:
passive luminosity evolution from the aging of the stellar populations,
and mass evolution due to dynamical effects such as mergers, 
tidal stripping and dynamical friction.
Cluster galaxies are known to consist of old stellar populations, with the
bulk of their star formation occurring at $z$$\gg$1. However,
they may not have been completely assembled by $z\sim 1$, and they 
may still undergo one or two significant mergers since then, or grow by gas 
accretion (van Dokkum {\it et al.}, 1999). 

The observed evolution of the luminosity function is a powerful 
probe for the assembly epoch of galaxies.
From a theoretical point of view, hierarchical models of galaxy formation 
based on semi--analytical prescriptions tend to predict a deficit of massive 
galaxies at $z\sim 1$, as a result of the ongoing mass assembly activity 
at lower redshift (Kauffmann \& Charlot, 1998).
More recent models, based on a flat $\Lambda$CDM cosmology
and with updated star formation, feedback and wind prescriptions
perform much better and agree with the observed number of massive galaxies 
up to $z=1.2-1.5$; however they still face problems with a deficit at higher
redshifts as well as with the colours of massive galaxies, underpredicting 
the number of red early type galaxies and of Extremely Red Objects. 
In the deep field, the observed evolution 
seems to lie in between the predictions of current hierarchical models
and those of the monolithic, pure luminosity evolution scenario 
(Somerville {\it et~al}.\ 2004; Stanford {\it et~al}.\ 2004).

For clusters instead, recent results from the observations of distant objects 
suggest 
that the evolution of the LF is best modelled by pure luminosity evolution:
Barger {\it et al.} (1996, 1998), De Propris {\it et al.} (1999), Kodama 
\& Bower (2003) found that, after correcting for this effect 
and for cosmological dimming, the galaxy mass function has actually evolved 
little over time; the K--band LF at $z \sim 1$ is consistent with pure 
luminosity evolution with constant stellar mass and early redshifts of star 
formation ($z_f \ga 2$).

\begin{figure}
\includegraphics[width=85mm]{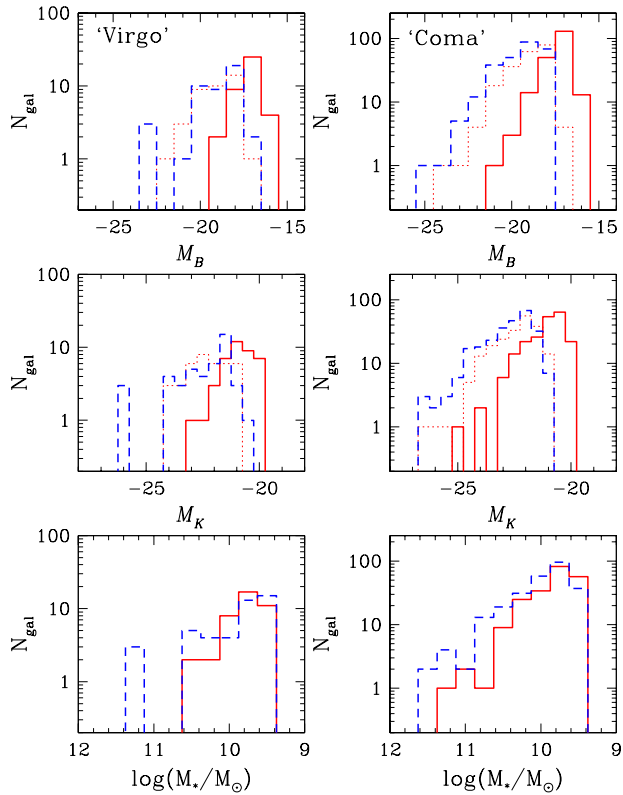}
\caption{{\it Top panels:} Luminosity functions in absolute number of
galaxies per {\it B}-band magnitude bin for the standard (AY-SW) ``Virgo'' 
and ``Coma'' simulations, at $z=0$ ({\it solid 
lines}) and $z=1$ ({\it dashed lines}); the dotted lines represents
the expected $z=1$ LF if pure passive evolution of the stellar populations is
applied to the $z=0$ LF. {\it Middle panels:} Same, for the K-band luminosity
functions.
{\it Bottom panels:} Mass function of cluster
galaxies at $z=0$ and $z=1$.}
\label{fig:LFz}
\end{figure}

In Fig.~\ref{fig:LFz} we show the evolution of the B and K band LFs 
of our ``Virgo'' and ``Coma'' clusters. We computed the (rest--frame) LF 
of the galaxies identified in the $z=1$ frame; this is shifted to brighter 
magnitudes with respect to the distribution at $z=0$, and
both luminosity evolution and dynamical mass evolution are seen to play a role
in our simulated clusters. In particular, 
since the bulk of the stars in our galaxies are formed at $z \gsim 2$ 
(see Paper~III and Fig.~\ref{fig:redsequence}, bottom panel),
luminosity dimming is an important effect.

We also computed the LF at $z=1$ as expected from pure passive evolution of
the $z=0$ LF, by considering the star particles in each of the galaxies
identified at $z=0$, and computing the luminosity they had back 
at a $\sim$7.5~Gyr younger age.
Such ``pure passive evolution'' LF is displayed as a dotted line
in Fig.~\ref{fig:LFz}, and it matches quite closely the actual LF in the $z=1$ 
frame (apart from the absence of a few bright galaxies, see below).
This indicates that luminosity dimming of the stellar populations
drives most of the LF evolution and the fading to fainter 
magnitudes from $z=1$ to $z=0$ for the simulated galaxy population; 
this is in line with the results of Kodama \& Bower (2003).

In the brightest luminosity bins of our LF, some additional,
dynamical effect seems to be required since more bright galaxies 
are identified at $z=1$ than expected from pure passive evolution;
especially in the case of the ``Coma'' simulation, where 263 galaxies ar found
at $z=1$ versus 212 at $z=0$. 
To assess mass evolution effects, we plot in the lower panels of
Fig.~\ref{fig:LFz} the mass function of (the stellar component of) cluster 
galaxies at $z=1$ and at $z=0$. The two mass functions are very
similar at the low mass end, but a few rather massive objects 
(above $M_* \sim 4 \times 10^{10}$~$\Msun$) present at $z=1$ 
``disappear'' at $z=0$. Inspection of the simulations show that these objects
have merged with the cD at $z$=0. 

As shown in \S~\ref{sect:resolution}, numerical ``over--merging'' onto 
the central cD cannot be the main reason for missing galaxies 
at the bright end of the simulated LF.
The lack of bright, massive galaxies other than the cD at $z$=0 
is probably largely affected by our group and cluster selection procedure, 
which picks out extremely relaxed objects at z=0,
as discussed in \S\ref{sect:lf}. 
All in all, ``Coma'' presents a less relaxed structure than ``Virgo'',  
as also inferred by the shapes of the LF at z=0 and z=1. Indeed even
for ``Virgo'' we do see
a deficiency of bright galaxies at z=0, but not at z=1, where a gap is
just due to poorer statistics of galaxy numbers.
Work is in progress however to analyze
a large sample of groups more randomly selected (hence including both
relaxed and unrelaxed object) to assess the bias induced by the selection
procedure (Sommer--Larsen, D'Onghia \& Romeo 2005, in preparation).

\subsection{Passive evolution and the Fundamental Plane}

In the previous section we computed the pure passive evolution dimming,
from $z=1$ to $z=0$, corresponding to the cluster galaxies identified at $z=0$.
Constraints on the passive evolution of ellipticals in clusters are derived
from the observed evolution of the Fundamental Plane (FP), indicating 
a dimming of about 1.2 magnitudes in the B band in the redshift range 
$z=1 \rightarrow 0$; assuming a Salpeter slope for the IMF, this implies 
a redshift of formation $z_{for} \geq 2.5$ for the stellar populations 
(Van Dokkum \& Stanford 2003; Wuyts {\it et al.}\ 2004; Renzini 2005;
Holden {\it et al.} 2005).

A direct comparison to FP constraints is hampered by the fact
that in our LF we miss the bright, massive elliptical galaxies defining
the observed FP; nevertheless, it is interesting to comment on the predicted 
passive evolution of our galaxies. 
In the table below we show the B magnitude evolution 
$\Delta M_B= M_B(z=0) - M_B(z=1)$ predicted by the SSPs in use, as a function 
of the assumed redshift of formation $z_{for}$ of the stellar population.
These values are largely independent of the SSP metallicity,
at least within a factor of 3--4 of the solar value (which is certainly
representative of the bright galaxies in our simulations).

\begin{center}
\begin{tabular}{c|c c}
$z_{for}$ & $\Delta M_B$ (Salp) & $\Delta M_B$ (AY) \\
\hline
     5    &        0.97         &      1.07         \\
     3    &        1.15         &      1.27         \\
    2.5   &        1.26         &      1.40         \\
    1.5   &        2.01         &      2.20         \\
\end{tabular}
\end{center}

\noindent
Since younger stellar populations dim faster, the lower the redshift of
formation, the faster the magnitude evolution between $z=1$ and $z=0$.
For the Salpeter IMF, our photometric code predicts that the observed dimming 
of $\sim$1.2 mag, corresponds to $z_{for} \geq$2.5 in agreement with the above 
mentioned studies. The rate of dimming depends however also on the slope
of the IMF, being faster for shallower IMFs; for an AY slope, 
the luminosity evolution is in fact slightly faster 
requiring $z_{for} \geq 3$ (see also Renzini 2005).
Taking into account progenitor bias, i.e.\ the fact the progenitors
of the youngest present--day early-type galaxies drop out of the sample
at high redshift, the actual magnitude evolution of ellipticals might be
underestimated by about 20\% (van Dokkum \& Franx 2001). In this case,
$\Delta M_B=1.4$~mag and $z_{for}=2.5$ (AY IMF) are still compatible
with observations.

The mean redshift of formation for the stars in our simulated galaxies 
is ~2.5 (Paper~III), which is on average compatible with FP constraints
but the scatter around the mean is large. Younger star particles, 
where present, dominate in the luminosity 
contribution at $z=1$ and induce a much faster overall evolution 
(cf.\ a stellar population formed at z=1.5 dims by more than 2 mag). 
This scatter in age makes some of our galaxies evolve faster than 
indicated by FP studies. 

\begin{figure}
\includegraphics[width=85mm]{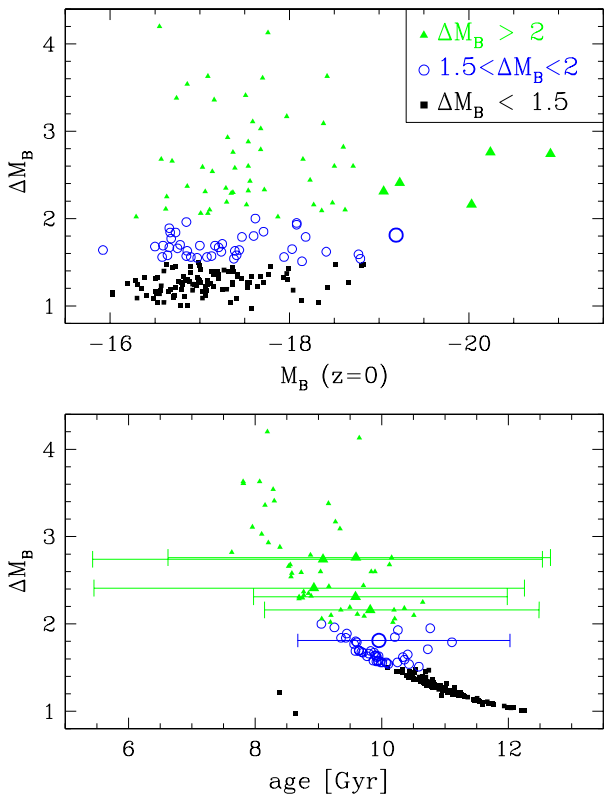}
\caption{Passive evolution in B magnitude between $z=1$ and $z=0$, predicted 
for the Coma cluster galaxies. $\Delta M_B = M_B(z=0) - M_B(z=1)$ is plotted
as a function of the present--day magnitudes at $z=0$ ({\it top panel}) 
and as a function of age ({\it bottom panel}). For the six brightest objects
(largw symbols) we also show the internal spread in stellar ages responsible
for the very large magnitude evolution (see text).}
\label{fig:evolMB}
\end{figure}

This is illustrated in Fig.~\ref{fig:evolMB}, where we show
the magnitude evolution $\Delta M_B=M_B(z=0)-M_B(z=1)$ for our simulated
``Coma'' galaxies as a function of 
their present--day magnitude and as a function of average 
stellar age. 
The age--dimming (i.e., $z_{for}-\Delta M_B$) relation is well defined in the
bottom panel and the FP constraint of 1.2-1.4 mag is met by objects older
than 10.5-11~Gyr.
The scatter above the lower envelope of the age--dimming relation
is due to internal age scatter within the galaxies. We analyze in particular
the six brightest objects in the simulation, with $M_B<-19$ (large symbols 
in the figure), which are more relevant for comparison to the bright 
spheroidals on the FP. These objects are 9--10 Gyr old (average age), i.e.\ 
$z_{for}$=1.5--2 which corresponds to $\Delta M_B=1.5-2$.
However, they also present a large internal scatter in stellar age, 
as shown by the errorbars in the bottom panel (stretching from the minimum to 
the maximum stellar age in each object). In particular, they contain stars 
as young as 8~Gyr, and some of them have minor tails of SF stretching 
to $z<1$, i.e.\ to ages younger than 7.5~Gyr.
The presence of these younger--than--average stars has negligible effects 
on the present--day appearance of these galaxies, since at $z=0$ 
all the stars are already quite ``old'' (ages $>5$~Gyr) --- see also 
Fig.~\ref{fig:redsequence}d showing that the luminosity age equals the
actual average age for these objects, while it would be significantly younger
if the younger stellar component were prominent. However, back at $z=1$ these
younger--than--average stars become very bright, young populations 
(ages $\leq$1~Gyr) with a major contribution to luminosity, so that
they ultimately drive the overall brightening of the galaxy when we trace
it back to to $z=1$.

The median of the luminosity evolution for all the Coma galaxy population
is 1.4~mag (in agreement with the average $z_{for}=2.5$), i.e.\ close to 
the FP constraint; however the large internal age scatter within the galaxies, 
especially the brighter ones, induces an overall predicted passive evolution 
for the LF more extreme than that (Fig.~\ref{fig:LFz}).

Higher resolution simulations can possibly help with this problem, by both 
anticipating the overall SF process (resolving smaller, hence denser, 
substructures) and reducing the internal poissonian noise in the stellar age 
distribution of the individual galaxies. However, the high resolution test
presented in \S~\ref{sect:resolution} does not show major differences 
in the LF down to $z=0.8$ (Fig.~\ref{fig:LFhrV}), suggesting that we have 
reached a resolution good enough to grasp the main galaxy properties.
We are probably then facing here a standard problem of hierarchical models
of galaxy formation, i.e.\ that they predict both slightly younger average
ages (Fig.~\ref{fig:redsequence}d) and a larger (internal) age scatter 
for more massive objects, both trends in contrast with observational evidence
(see Section~\ref{sect:SFH} and references therein). 
 
We notice however that the faint end of the present--day 
colour--magnitude relation of E+S0 cluster galaxies seems to be largely
populated by objects which reddened onto the Red Sequence only recently,
while at high $z$ they were much bluer; an effect that is usually
associated
to the spiral $\longrightarrow$ S0 transformation and to the
Butcher--Oemler
effect from intermediate redshift to $z=0$ (e.g.\ De Lucia {\it et~al.}\
2004;
see also Section~\ref{sect:SFH} and references therein). These objects are
certainly not the same probed by the FP at high redshift,
which relates to the most massive ellipticals. 
Also, Holden {\it et~al.}\ (2005) find a large scatter in mass--to--light
ratio and hence magnitude evolution for the least massive ellipticals
selected at high $z$, with masses around $10^{11}~\Msun$ (comparable
to the most massive objects in our final galaxy population,
Fig.~\ref{fig:LFz}).
Henceforth, it is possible that
our simulations miss in fact exactly the massive spheroidals at the bright
end of the LF, which 
are most meaningful for FP constraints.
The issue needs then to be revisited once the problem of properly
populating 
the bright end of the LF is solved for simulated clusters.

\section{The Red Sequence}
\label{sect:RS}

\begin{figure}
\includegraphics[width=85mm]{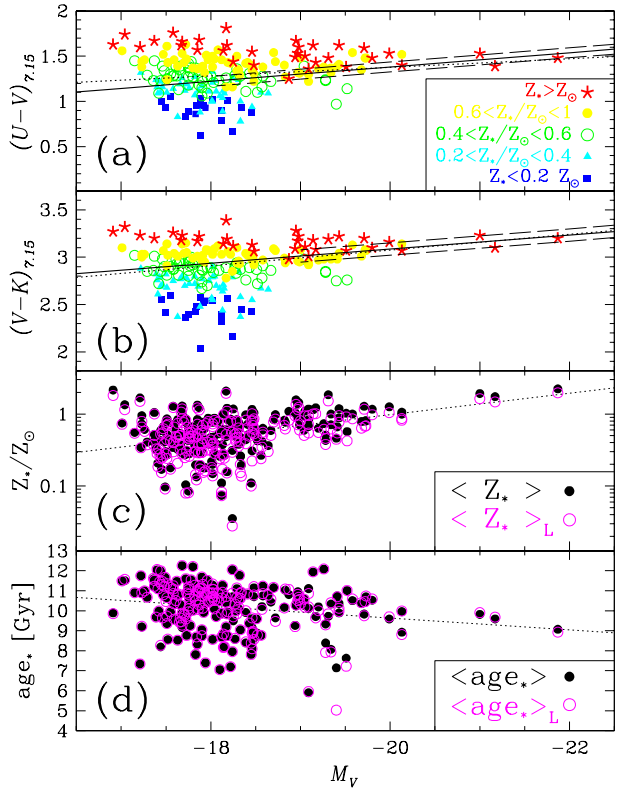}
\caption{{\bf (a--b)} Colour--magnitude relation for galaxies in the
simulated ``Coma'' cluster (data points and {\it dotted} linear fit line),
compared to the observed relation with its scatter 
({\it solid} and {\it long--dashed} lines, from Bower {\it et~al.} 1992; 
{\it short--dashed} lines, from Terlevich {\it et~al.}\ 2001).
The extension of the dashed lines
indicates the magnitude range probed observationally, while the average
fit relation (solid line) has been extrapolated to lower magnitudes.
{\bf (c)} Metallicity--luminosity relation for the ``Coma'' galaxies 
(full symbols for mass--averaged stellar metallicity, open symbols for 
luminosity--weighted metallicity); the dotted line is a linear fit.
{\bf (d)}: Age--luminosity relation for the ``Coma'' galaxies (mass--average
and luminosity--weighted stellar ages); the dotted line is a linear fit.}
\label{fig:redsequence}
\end{figure}

The light and stellar mass in clusters of galaxies is dominated by bright,
massive ellipticals (Abell 1962, 1965). These are known
to form a tight colour-magnitude relation, or Red Sequence (Bower, Lucey \& 
Ellis 1992ab; Gladders {\it et~al.}\ 1998; Andreon 2003; Hogg {\it et al.}
2004; McIntosh {\it et al.} 2005). In Fig.~\ref{fig:redsequence}ab
we compare the colour--magnitude relation for our ``Coma'' cluster galaxies,
excluding the cD, to the observed Red Sequence of Coma from
Bower {\it et~al.}\ (1992) and Terlevich, Caldwell \& Bower (2001),
assuming a distance modulus to Coma of 35.1, i.e.\ 
$H_0 \sim$70~km~sec$^{-1}$~Mpc$^{-1}$ (Baum {\it et~al.}\ 1997).
The slope of the observed Red Sequence is sensitive to aperture effects:
fixed--aperture observations are often used, measuring a smaller fraction
of the total light in larger objects than in smaller ones; combined with
colours gradients, this introduces a bias in the sense of measuring
systematically redder colours in larger galaxies, i.e.\ a
steepening of the actual colour--magnitude relation. When
the analysis is instead performed on an area that scales with the size of
the galaxy, e.g.\ within the effective radius, or within a given
isophotal/photometric radius, the colour--magnitude relation is in fact
flatter (Terlevich {\it et~al.}\ 2001; Scodeggio 2001; and references therein).
For Coma and Virgo, Bower {\it et~al.}\ measured colours within a fixed
aperture of 11'' and 60'' respectively, corresponding to a fixed radius  
of $5 h^{-1} \sim 7.15$~kpc. Henceforth for a proper comparison, we computed
and plotted in Fig.~\ref{fig:redsequence} the colours for our simulated
galaxies within a projected radius of 7.15~kpc. The magnitude in abscissa
are instead total magnitudes, as adopted by Bower {\it et~al.}

The solid lines in panels (a,b) indicate the observed slope and location
of the Red Sequence; the dotted line is a least square fit to our data.
There is overall good agreement, although our Red Sequence is not as extended
as the observed one (which reaches magnitudes as bright as $M_V=-23$), 
due to the lack of simulated galaxies at the bright end of the luminosity 
function discussed in the previous section. 
Our slope for the (U-V) vs.\ V relation (Fig.~\ref{fig:redsequence}a), 
is slightly flatter than the observed one, but the colours of the three  
brightest objects are in good agreement with observations, even within
the observed small scatter. These objects are the most significant ones
if we remember that the observed slope is defined only for $M_V < -19$, 
extended down to $M_V < -18$ by Terlevich {\it et~al.} The slope of the   
simulated (V-K) vs.\ V relation (Fig.~\ref{fig:redsequence}b) is in perfect
agreement with the observed one.

Although we formally obtain the right slopes, the simulated Red Sequence also
displays a much larger scatter than the observed Coma Red Sequence, which 
is very narrow down to $M_V \sim -18$ (dashed and dotted lines in 
Fig.~\ref{fig:redsequence}ab; see also Fig.~2a in Gladders {\it et~al.}, 1998).
Such large scatter is partly due to 
Poissonian noise from the small number of star particles in the low 
luminosity galaxies. Besides, while the observed colour--magnitude relation
refers to early type galaxies (E/S0), we did not attempt any morphological 
classification and {\it a priori} selection of our simulated galaxies
due to limited resolution.

The colour--magnitude relation is classically interpreted as a 
mass--metallicity relation (Kodama \& Arimoto 1997). In our simulations, 
the bulk of the stars in galaxies are formed at $z\ga$2 (see also Paper~III), 
hence age trends among galaxies are only mild (Fig.~\ref{fig:redsequence}d) 
and the colour--magnitude relation is mostly a metallicity effect
(Fig.~\ref{fig:redsequence}c). The metallicity--luminosity relation 
for our cluster galaxies is displayed Fig.~\ref{fig:redsequence}c, 
both in terms of mass-weighted metallicity and of (V--band) 
luminosity--weighted 
metallicity. The latter is systematically slightly lower than the actual 
mass-averaged stellar metallicity, as expected since more metal--poor 
populations tend to be brighter, skewing the luminosity--weighted metallicity 
to lower values (Greggio 1997); the trend with galaxy luminosity is however
maintained. Although metal-rich objects seem to exist at all 
luminosities, the fraction of metal--poor galaxies (as well as the scatter in
metallicity) increases with decreasing galaxy luminosity; the average 
stellar metallicity decreases for fainter galaxies (dotted line), which 
ultimately drives the simulated colour--magnitude relation.

Fig.~\ref{fig:redsequence}d shows the average stellar ages
for the galaxies. All the galaxies consist of old stellar populations,
with average ages between 7 and 12 Gyrs. The brighter galaxies ($M_V < -20$) 
are essentially coeval, 9--10~Gyr old. 
The luminosity--weighted ages are generally very close to the mass
averaged ages, which implies a small scatter of stellar ages within the
individual galaxies: extended tails or recent episodes of star formation, 
would strongly skew the luminosity--weighted estimate to younger ages.
The scatter in age among the galaxies is around 30\%, with a mild trend 
of decreasing age with increasing luminosity (dotted line), but the effect 
is much smaller that the systematic variation in metallicity (a factor of 10,
Fig.~\ref{fig:redsequence}c). This confirms that the color--magnitude 
relation of our simulated galaxies is driven by a mass-metallicity relation,
in agreement with common wisdom. The mild age trend, which tends to act in the
opposite direction (making fainter galaxies redder) is probably responsible of
the fact that our colour-magnitude relation is not as steep as
the observed one.

At all magnitudes, even the faintest ones, we find some simulated
galaxies with large (super--solar) metallicities.
For dwarf ellipticals, a very large spread in metallicities and colours
with a tail of red, metal rich objects,
is in fact observed but at fainter magnitudes than those probed
here ($M_B>-15$, Conselice {\it et~al.}\ 2003).
Metal rich dwarf ellipticals in clusters may possibly originate from
larger (and hence more metal rich) galaxies 
that have suffered considerable tidal stripping in the cluster 
potential. To test
this hypothesis we traced the evolution of the nine galaxies with
$Z_* > Z_{\odot}$ and $M_V > -18$, shown in the top panel of 
Fig.~8, back in time. Two of the galaxies were found to contain 
a significantly larger mass of stars at $z \sim 1.5-2.5$ 
(when these galaxy reach their maximum stellar masses),  whereas the
stellar masses of the other seven were found not to change much 
since their "birth".
Therefore, in our simulations, stripping of originally larger galaxies
is one possible, but not predominant, channel to form dwarf 
galaxies of high metallicity.

\begin{figure}
\includegraphics[width=85mm]{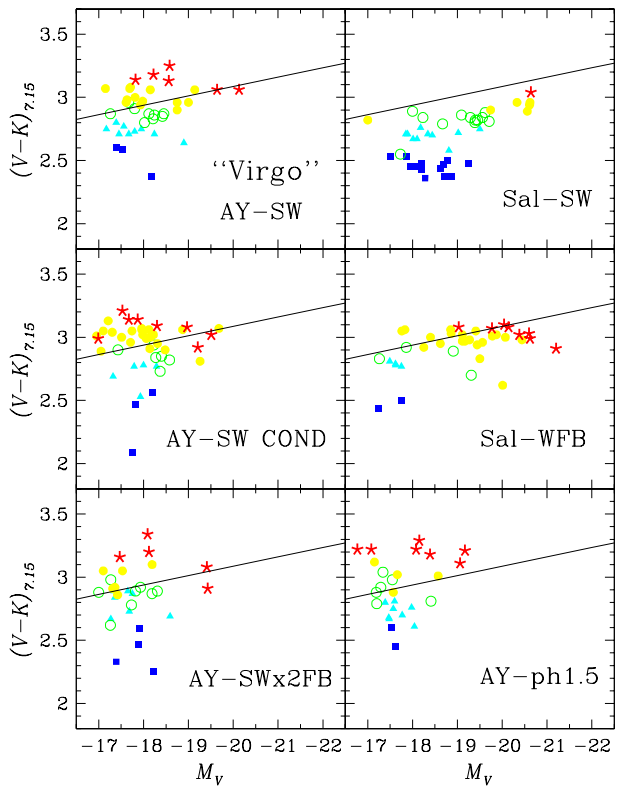}
\caption{Red Sequence for our simulated ``Virgo'' cluster with different
physical input. AY: Arimoto \& Yoshii IMF; Sal: Salpeter IMF; SW:
``standard'' feedback prescription; SWx2: two times stronger feedback
efficiency; WFB: weak feedback (strong feedback active only at early times);
COND: thermal conduction included; ph: energy preheating at $z=3$. 
Symbols follow the metallicity coding 
of Fig.~\protect{\ref{fig:redsequence}}.} 
\label{fig:rsVirgo}
\end{figure}

In Fig.~\ref{fig:rsVirgo} we assess the dependence of the simulated Red 
Sequence on the parameters of the simulation, by considering the ``Virgo'' 
cluster computed with different physical prescriptions (see 
Section~\ref{sect:simulations} and Table~\ref{tab:prescriptions}). 
For all the AY simulations the galaxies scatter about the observed Red 
Sequence, and their average location in the colour--magnitude diagram 
appears to be robust with respect to the inclusion of thermal 
conduction or preheating, and also to enhanced feedback efficiency
(AY-SWx2; we do not display the case AY-SWx4 since the total number of 
galaxies and their average mass are quite small --- see Fig.~\ref{fig:LFV} --- 
but they still scatter around the observed line). The scatter is however 
very large and the magnitude range of the galaxies formed quite limited 
(missing the bright end of the LF), so that one cannot make meaningful 
comparisons to the observed slope of the Red Sequence.

Significantly different is the case of the Salpeter IMF simulations. 
The scatter is much reduced and the slope of the observed colour--magnitude 
relation is well reproduced; however,
the simulated galaxies are offset to the blue of the observed Red Sequence.
Let us discuss the latter problem first.
In the standard SuperWind case (Sal-SW) in particular, the stellar populations 
are too metal poor, hence too blue, to match the observed 
Red Sequence. Notice that the blue colours
are not due to younger stellar ages, since the star formation history in the
Virgo cluster is very similar for the Salpeter and the AY simulations
(Fig.~\ref{fig:SFRz}). In the weak feedback scenario, where effective
supernova energy injection is limited to the early epochs (Sal-WFB) the
simulated Red Sequence falls closer to the observed one, though still on the 
blue side. With minor feedback the galaxies
retain a much larger fraction of the metals they produce, thereby the stellar
metallicities can grow larger; however, the price is a predicted enrichment 
of the ICM far below the observed levels for the Sal-WFB simulations
(Table~\ref{tab:prescriptions} and Paper~I).  
Henceforth, top--heavier IMFs than the Salpeter IMF are
needed, not only for the sake of the ICM enrichment but also to reproduce
the observed colours and metallicities of the stellar populations. In fact
the Salpeter IMF does not produce enough metals, and/or locks too much mass
in stars, to account for the observed metallicities in the ICM and in
cluster galaxies at the same time (Portinari {\it et~al.}\ 2004, and Paper I).

As to the increased dispersion in galaxy properties for the AY vs.\ Salpeter 
simulations, this is likely induced by the much stronger feedback 
(a combination of the SW prescription with a top--heavy IMF).
As a consequence, the Red Sequence in the AY simulations is much less tight,
with no well-defined slope, although the average colours of the galaxies
agree with observations much better than in the Salpeter case. If the 
dispersion is mostly due to numerical effects, because of the small number 
of particles in the individual
galaxies, higher resolution simulations should show a reduced
scatter --- something to be tested in future work. 

In summary, the zero--point of the simulated Red Sequence 
appears to be a robust prediction of the simulations, quite unaffected 
by the adopted physical prescriptions other than the chosen IMF, which sets
the typical stellar metallicity attainable in cluster galaxies.
The observed slope of the colour--magnitude relation is well reproduced 
in the Salpeter simulations (plagued however by an offset to too blue colours
and low metallicities); for the AY simulations the scatter is much
larger but the average colours are better reproduced. 

\section{Conclusions}
In this paper we have presented for the first time
and analyzed the properties of the galaxy population in clusters,
as predicted from full {\it ab initio} cosmological + hydrodynamical
simulations.

Our results are based on cosmological
simulations of galaxy clusters including self-consistently metal-dependent
atomic radiative cooling, star formation, supernova and (optionally) AGN 
driven galactic 
super-winds, non-instantaneous chemical evolution, effects of a
meta-galactic, redshift dependent UV field and thermal conduction.
In relation to modelling the properties of cluster galaxies this is an 
important step forward with respect to 
previous theoretical works on the subject, e.g. based on semi--analytical
recipes super--imposed on N--body only simulations.

The global star formation rates of the ``Virgo'' and ``Coma'' cluster 
galaxies are found to 
decrease very significantly with time from redshift $z$=2 to 0, in agreement
with what is inferred from observations of the inner parts of rich clusters
(e.g., Kodama \& Bower 2001).

We have determined galaxy luminosity functions for the ``Virgo'' and 
``Coma'' clusters in the $B, V, R, H$ and $K$ bands; the comparison to
observed galaxy luminosity functions 
reveals a deficiency of bright galaxies ($M_{\rm{B}}$$\la$--20). 
We carried out a test simulation of ``Virgo'' at eight times
higher mass resolution and two times higher force resolution;
the results of this test, still running at the present, 
indicate that the above mentioned deficiency of bright galaxies 
is {\it not} due to ``over--merging''; higher resolution simulations of
``Coma'' clusters are in progress as well to further 
check this point. 
From a suite of simulations for the ``Virgo'' cluster with different
input physics, we find that the deficiency of bright galaxies becomes
less prominent with decreasing super-wind strength, in particular
for models invoking a Salpeter IMF and only early feedback; in fact more
mass can be accumulated in stars and galaxies, with low feedback. Such models,
however, present various drawbacks: the cold fraction is too high and the
metal production is too low, as seen in the too blue colours of the
galaxies and/or in the low metallicity of the ICM, which can hardly
be enriched to the observed level of about 1/3 of solar abundance; the latter
point is discussed in detail in Paper~I, but it also follows
from more general arguments (Portinari {\it et~al.} 2004).
The bright galaxy deficiency might be explained as a selection effect, 
in the sense that
we have selected cluster haloes for the TreeSPH re-simulations, which are
``too relaxed'' compared to an average cluster halo,
so that the brightest galaxies have by now merged into the central cD
by dynamical friction; we shall return to
this in a forthcoming paper.

The redshift evolution of
the luminosity functions from redshift $z$=1 to 0 is mainly driven by 
passive luminosity evolution of the stellar populations,
but also by the above mentioned merging of bright galaxies into the cD.

The slope of the colour--magnitude relation of the simulated galaxies
is in good agreement with the observed one, however the scatter is larger
than observed, partly due to poissoinian noise within the fainter galaxies
which are formed by small numbers of star particles.
Such internal dispersion in stellar ages is also responsible for
a luminosity dimming between $z=1$ and $z=0$,
faster than indicated by the observed evolution of the Fundamental Plane.
The typical average galaxy colours are best matched
when adopting a top-heavy IMF (as originally suggested by Arimoto \& Yoshii
1987), while Salpeter IMF simulations yield too blue colours.
Moreover we find that the average metallicity of the simulated
galaxies increases with luminosity, and that the
brightest galaxies are essentially coeval. Hence, the colour--magnitude
relation results from metallicity rather than age effects, as concluded
by Kodama \& Arimoto (1997) on the basis of its observed evolution.
\section*{Acknowledgments}
We thank S.\ Andreon, T.\ Kodama, S. Leccia, Y.-T. Lin for fruitful
discussions, and/or
for having provided us with their observational data.\\
\indent
All computations presented in this paper have been performed on the IBM SP4 
system at the Danish Centre for Scientific Computing (DCSC).

This work was supported by Danmarks Grundforskningsfond through its 
contribution to the establishment of the Theoretical Astrophysics Centre (TAC),
by the Villum Kann Rasmussen Foundation 
and by the Academy of Finland (grant nr.~206055).
\bibliographystyle{aa}

\end{document}